\documentclass[12pt]{article}
\usepackage{amsmath, amsthm, amssymb, setspace, mathrsfs, color, verbatim, graphicx, hyperref, amsfonts, bm, rotating, lscape, multicol, multirow, natbib, dsfont, array} 
\usepackage[left=0.7in,right=1.0in,top=0.7in,bottom=1.0in,nohead]{geometry}
\usepackage[mathscr]{euscript}
\def\bse{\begin{eqnarray*}}
\def\ese{\end{eqnarray*}}
\def\be{\begin{eqnarray}}
\def\ee{\end{eqnarray}}

\newcolumntype{C}[1]{>{\centering\let\newline\\\arraybackslash\hspace{0pt}}m{#1}}
\newcolumntype{L}[1]{>{\raggedright\let\newline\\\arraybackslash\hspace{0pt}}m{#1}}
\newcolumntype{R}[1]{>{\raggedleft\let\newline\\\arraybackslash\hspace{0pt}}m{#1}}

\begin{document} 

\title{A Spatio-Temporal Point Process Model \\ for Ambulance Demand}
\author{Zhengyi Zhou, David S. Matteson, Dawn B. Woodard, \\ Shane G. Henderson and Athanasios C. Micheas\footnote{
Zhou is a PhD Candidate at the 
Center for Applied Mathematics,
Cornell University,
657 Rhodes Hall,
Ithaca, NY 14853
(E-mail: \href{mailto:zz254@cornell.edu}{zz254@cornell.edu}; Webpage: \url{http://www.cam.cornell.edu/\~zz254/}).
Matteson is an Assistant Professor at the 
Department of Statistical Science,
Cornell University,
1196 Comstock Hall,
Ithaca, NY 14853
(Email: \href{mailto:matteson@cornell.edu}{matteson@cornell.edu}; Webpage: \url{http://www.stat.cornell.edu/\~matteson/}).
Woodard is an Assistant Professor at the 
School of Operations Research and Information Engineering,
Cornell University,
228 Rhodes Hall,
Ithaca, NY 14853
(Webpage: \url{http://people.orie.cornell.edu/woodard/}).
Henderson is a Professor at the 
School of Operations Research and Information Engineering,
Cornell University,
230 Rhodes Hall,
Ithaca, NY 14853
(Webpage: \url{http://people.orie.cornell.edu/shane/}).
Micheas is an Associate Professor at the 
Department of Statistics,
134G Middlebush Hall,
University of Missouri-Columbia,
Columbia, MO 65203
(Webpage: \url{http://www.stat.missouri.edu/\~amicheas/}).
}}
\date{\today}
\maketitle


\begin{abstract}
Ambulance demand estimation at fine time and location scales is critical for fleet management and dynamic deployment. We are motivated by the problem of estimating the spatial distribution of ambulance demand in Toronto, Canada, as it changes over discrete 2-hour intervals. This large-scale dataset is sparse at the desired temporal resolutions and exhibits location-specific serial dependence, daily and weekly seasonality. We address these challenges by introducing a novel characterization of time-varying Gaussian mixture models. We fix the mixture component distributions across all time periods to overcome data sparsity and accurately describe Toronto's spatial structure, while representing the complex spatio-temporal dynamics through  time-varying mixture weights. We constrain the mixture weights to capture weekly seasonality, and apply a conditionally autoregressive prior on the mixture weights of each component to represent location-specific short-term serial dependence and daily seasonality. While estimation may be performed using a fixed number of mixture components, we also extend to estimate the number of components using birth-and-death Markov chain Monte Carlo. The proposed model is shown to give higher statistical predictive accuracy and to reduce the error in predicting EMS operational performance by as much as two-thirds compared to a typical industry practice.

\end{abstract}

\par\vfill\noindent
\par\medskip\noindent
{\bf Short title: Spatio-Temporal Ambulance Demand}

\clearpage\pagebreak\newpage
\newlength{\gnat}
\setlength{\gnat}{24pt} 
\baselineskip=\gnat

\section{Introduction}\label{introduction}

This article describes an efficient and flexible method to model a time series of spatial densities. The motivating application is estimating ambulance demand over space and time in Toronto, Canada. Emergency medical service (EMS) managers need accurate demand estimates to minimize response times to emergencies and keep operational costs low. These demand estimates are typically required at least for every four-hour work shift, and at very fine spatial resolutions for fine-grained dynamic deployment planning. We propose a method to model ambulance demand continuously on the spatial domain, as it changes over every two-hour interval. 

Several studies have modeled aggregate ambulance demand as a temporal process. \cite{Channouf:2007} use autoregressive moving-average models for daily demand in Calgary, Canada, and estimate hourly demand conditional on the daily total. 
\cite{Matteson:2011} directly model hourly call arrival rates in Toronto, Canada by combining a dynamic latent factor structure with integer time series models. 
Other aggregate demand studies have also considered singular spectrum analysis \citep{Vile:2012}, fixed-effects, mixed-effects and bivariate models \citep{Aldor:2009, Ibrahim:2013}, Bayesian multiplicative models \citep{Weinberg:2007} and Singular Value Decomposition \citep{Shena:2008, Shenb:2008}.
While these temporal estimates inform staffing and fleet size, spatio-temporal demand estimates are critical for selection of base locations and for dynamic deployment planning, but have received far less attention. 
Current industry practice for spatial-temporal demand forecasting usually uses a simple averaging technique over a discretized spatial and temporal domain; the demand estimate for a small spatial region in a particular time period is taken to be the average of several historical demand values for the same region during corresponding periods in previous weeks or years. Averages of so few data points can produce noisy estimates, and can vary greatly with changes in the discretization.  \cite{Setzler:2009} use artificial neural networks (ANN) on discretized spatial and temporal domains, and compare it to industry practice. ANN is superior at low spatial granularity, but both methods produce noisy results at high spatial resolutions. The proposed method is suitable for estimation on very fine scales in time and space. 

\begin{figure}[ht]
	\centering
\includegraphics[width=1.0\textwidth]{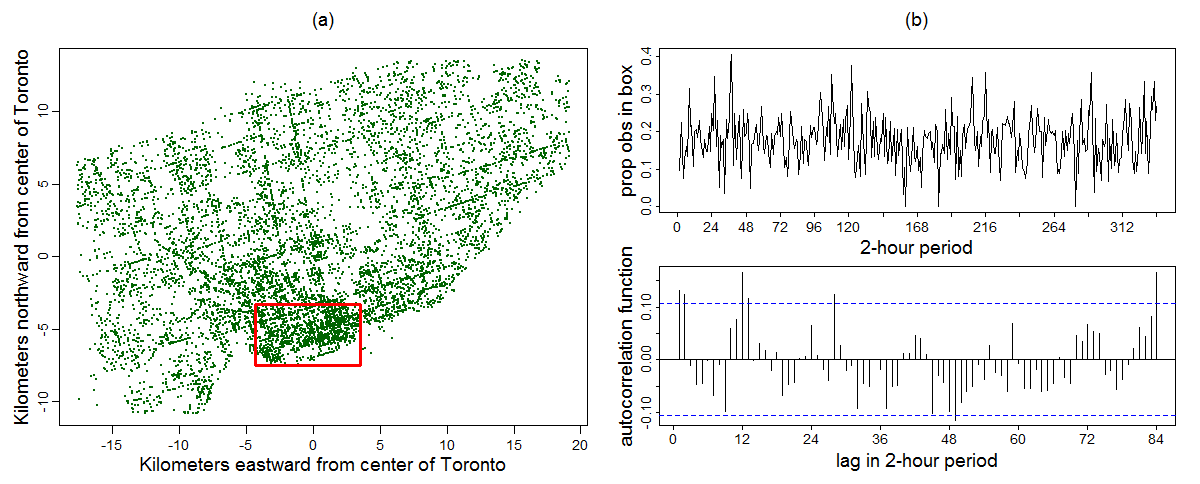}
	\caption{(a) all $15,393$ observations in the training data (February 2007), with downtown subregion outlined by a rectangle; (b) time series (top) and autocorrelation function (bottom) of the proportions of observations arising from the rectangle across two-hour periods. Weekly and daily seasonality, and low-order autocorrelation are observed.}
	\label{data}
\end{figure}

We use data from Toronto Emergency Medical Services, for February 2007, and evaluate out-of-sample performance on data from March 2007 and February 2008. The data consist of $45,730$ priority emergency events received by Toronto EMS for which an ambulance was dispatched. Each record contains the time and the location to which the ambulance was dispatched.  
Figure \ref{data} shows all observations from the training data (February 2007) and explores some characteristics of downtown Toronto. For each two-hour period, we compute the proportion of observations that arise from the downtown region (outlined by the rectangle) out of all observations from that period. We analyze the autocorrelation of the time series of proportions in Figure \ref{data}(b).  We observe evidence for weekly (84 time periods) and daily (12 time periods) seasonality as well as low-order autocorrelation (dashed lines represent approximate 95\% point-wise confidence intervals). Upon analyzing other localities, we consistently found weekly seasonality, but daily seasonality and low-order autocorrelation tend to be stronger at locations such as downtown or dense residential regions, and weaker at others such as dispersed residential areas or large parks. 

Existing approaches to estimate spatial or spatio-temporal densities of point processes do not fully address the additional challenges in EMS analysis. Spatial point processes have frequently been modeled using non-homogeneous Poisson processes (NHPP) \citep{Diggle:2003, Moller:2004, Illian:2008}. In particular, Bayesian semiparametric mixture modeling has been proposed to account for heterogeneity in the spatial intensity function. Examples include Dirichlet processes with beta or Gaussian densities \citep{Kottas:2007,Ji:2009}, and finite Gaussian mixture models with a fixed number of components \citep{Chakraborty:2010}. However, EMS data is sparse at the desired temporal granularity for estimation in this industry; the average number of observations in each two-hour period is only 45. This makes it difficult to estimate an accurate spatial structure at each time period.

Recently, dependent Dirichlet processes have been proposed to model correlated spatial densities across discrete time \citep{Taddy:2010, Ding:2012, Taddy:2012}. These methods allow the stick-breaking weights of the Dirichlet process to evolve in an autoregressive manner, but necessitates a simple first-order dependence structure common to all components. For EMS applications, it is essential to capture a much more complex set of temporal dynamics, including short-term serial dependence as well as daily and weekly seasonalities. Moreover, some of these dynamics vary from location to location. To consider only the first-order dependence, and enforce it across the entire spatial domain is very limiting. On the other hand, extending the dependent Dirichlet processes to include higher-order serial dependence and multiple seasonalities is not straightforward. It is also not easy to make these dynamics location-specific. Discretizing the spatial domain into sub-regions and imposing a different autoregressive parameter on each region would add substantial computational complexity, and is sensitive to spatial partitioning.

To address the stated modeling aims, we propose a novel specification of a time-varying finite mixture model. We assume a common set of mixture components across time periods, to promote effective learning of the spatial structure across time, and to overcome sparsity within each period. We allow the mixture weights to vary over time, capturing temporal patterns and dynamics in the spatial density by imposing seasonal constraints and applying autoregressive priors on the mixture weights.
The number of mixture components may be fixed or estimated via birth-and-death Markov chain Monte Carlo \citep{Stephens:2000}. 
We compare the proposed method with a current industry practice, as well as a proposed extension of this practice. The proposed method is shown to have highest statistical predictive accuracy, as well as the least error in measuring operational performance.


We define the general setting and propose a spatio-temporal mixture model using a fixed number of components in Section \ref{3models}. 
We extend the mixture model to estimate the number of components in Section \ref{nocomponents}. 
We show the results of estimating ambulance demand in Toronto in Section \ref{emp}, and assess the performance and validity of the proposed approach in Section \ref{perf}. Section \ref{conclusions} concludes.

\section{Spatio-Temporal Finite Mixture Modeling}\label{3models}

We investigate Toronto's ambulance demand on a continuous spatial domain $\mathcal{S}\subseteq \mathds{R}^2$ and a discretized temporal domain $\mathcal{T}=\{1,2, \ldots, T\}$ of two-hour intervals ($T=336$ for 28 days in February 2007). 
The proposed method trivially extends to other spatial domains. 

\indent Let $\boldsymbol{s}_{t,i}$ denote the spatial location of the $i$th ambulance demand event occurring in the $t$th time period, for $i \in \{1,\dots, n_t\}$. 
We assume that the set of spatial locations in each time period independently follows a non-homogeneous Poisson point process over $\mathcal{S}$ with positive integrable intensity function $\lambda_t$.
The intensity function for each period $t$ can be decomposed as $\lambda_t(\boldsymbol{s})=\delta_t f_t(\boldsymbol{s})$ for $\boldsymbol{s}\in \mathcal{S}$,
in which $\delta_t = \int_{\mathcal{S}} \lambda_t(\boldsymbol{s})\, d\boldsymbol{s}$ is the aggregate demand intensity, or total call volume, for period $t$, and $f_t(\boldsymbol{s})$ is the spatial density of demand in period $t$, i.e., $f_t(\boldsymbol{s})>0$ for $\boldsymbol{s} \in \mathcal{S}$ and $\int_{\mathcal{S}}f_t(\boldsymbol{s})\,d\boldsymbol{s}=1$. Then we have $n_t | \lambda_t \sim \mbox{Poisson}(\delta_t)$ and $\boldsymbol{s}_{t,i} | \lambda_t, n_t \stackrel{iid}{\sim} f_t(\boldsymbol{s})$ for $i \in \{1,\ldots,n_t\}$.

Many prior studies propose sophisticated methods for estimating $\{ \delta_t \}$.
Here, we focus on estimating $\{f_t(\boldsymbol{s})\}$, which has received little consideration in the literature.
The proposed model is constructed in three steps. We first introduce in Section \ref{model1} the general framework of mixture models with common component distributions across time. We add constraints on the mixture weights in Section \ref{model2} to describe weekly seasonality. We also place autoregressive priors on the mixture weights to capture location-specific dependencies in Section \ref{shrinkage}. Finally, the computational methods are described in Section \ref{compu}. For now, we fix the number of mixture components $K$; estimation of $K$ is incorporated in Section \ref{nocomponents}. 

\vspace{-2pt}
\subsection{A Spatio-Temporal Gaussian Mixture Model} \label{model1}

\vspace{-10pt}
We consider a bivariate Gaussian mixture model in which the component distributions are common through time, while mixture weights change over time. Fixing the component distributions allows for information sharing across time to build an accurate spatial structure, because each time period typically has few observations. It is also natural in this application, which has established hotspots such as downtown, residential areas and central traffic routes. Letting the mixture weights vary across time enables us to capture dynamics in population movements and actions at different locations and times. The proposed methods can be trivially extended to other distributional choices such as a mixture of bivariate Student's $t$ distributions. For any $t\in \mathcal{T}$, we model the spatial distribution by a $K$-component Gaussian mixture
\vspace{-10pt}
\begin{equation} \label{gmodel1}
f_t(\boldsymbol{s};\{p_{t,j}\}, \{\boldsymbol\mu_{j}\}, \{\boldsymbol\Sigma_{j}\}) = \sum_{j = 1}^{K}p_{t,j}\,\phi(\boldsymbol{s};\boldsymbol\mu_{j},\boldsymbol\Sigma_{j}), \;\; \forall \; \boldsymbol{s} \in \mathcal{S},
\end{equation}
\vspace{-2pt}
\noindent in which $\phi$ is the bivariate Gaussian density, with mean $\boldsymbol \mu_j$ and covariance matrix $\boldsymbol\Sigma_j$, for $j \in \{1,\ldots,K\}$.  
The mixture weights are $\{p_{t,j}\}$, satisfying $p_{t,j} \geq 0$ and $\sum_{j=1}^K p_{t,j} = 1$ for all $t$ and $j$.
In this specification, the number of components $K$ is assumed fixed across time. However, components can be period-specific when their weights are $0$ in all other periods.
The component means and covariances are also the same in all time periods; only the mixture weights are time-dependent. 
If a spatio-temporal covariate density is available, for example population density, it 
 may be added as an additional mixture component.
 Let $g_t(\boldsymbol{s})$ denote a (time-varying) spatial density and let
$f_t(\boldsymbol{s}; \{p_{t,j}\}, \{\boldsymbol\mu_{j}\}, \{\boldsymbol\Sigma_{j}\})=p_{t,0}g_t(\boldsymbol{s}) + (1-p_{t,0})\sum_{j = 1}^{K}p_{t,j}\,\phi(\boldsymbol{s};\boldsymbol\mu_{j},\boldsymbol\Sigma_{j}), 
$ 
in which 
$p_{t,0} \in [0,1]$ is a time-varying probability that ambulance demand arises directly from the covariate density. 
As such, in time period $t$, demand arises from component $j$ of the mixture  with probability $(1-p_{t,0})p_{t,j} \in [0,1]$, for $j\in\{1,\ldots,K\}$. 
The relative importance of a covariate is measured by $p_{t,0}$. However, such a covariate was not available in this application.

\vspace{-2pt}
\subsection{Modeling Seasonality with Constraints} \label{model2}

We observe weekly seasonality in ambulance demand across the spatial domain (Section \ref{introduction}).
We represent this weekly seasonality by constraining all time periods with the same position within a week (e.g., all periods corresponding to Monday 8-10am) to have common mixture weights.

Let $B \in \mathds{N}$ ($B \ll T$) denote a time block, corresponding to the desired cycle length. In this application, $B = 84$, the number of 2-hour periods in a week. Each $t\in \mathcal{T}$ is matched to the value of $b\in \{1,\ldots,B\}$ such that $b \; \mbox{mod} \; B = t \; \mbox{mod} \; B$. We modify Equation (\ref{gmodel1}) to 
\begin{equation} \label{model2f}
f_t(\boldsymbol{s}; \{p_{t,j}\}, \{\boldsymbol\mu_{j}\}, \{\boldsymbol\Sigma_{j}\})=f_b(\boldsymbol{s}; \{p_{b,j}\}, \{\boldsymbol\mu_{j}\}, \{\boldsymbol\Sigma_{j}\})=\sum_{j = 1}^{K}p_{b,j}\,\phi(\boldsymbol{s};\boldsymbol\mu_{j},\boldsymbol\Sigma_{j}),
\end{equation}
\noindent so that all periods with the same position within the cycle have the same set of mixture weights.

The usefulness of such constraints on mixture weights is not limited to representing seasonality. We can also exempt special times, such as holidays, from seasonality constraints, or combine consecutive time periods with similar characteristics, such as rush hours or midnight hours.

\subsection{Autoregressive Priors} \label{shrinkage}

We also observe that EMS demand exhibits low-order serial dependence and daily seasonality whose strengths vary with locations (Section \ref{introduction}).  
We may capture this in the proposed mixture model by placing a separate conditionally autoregressive (CAR) prior on each series of mixture weights, i.e., $\{p_{b,j}\}_{b=1}^B$ for each $j$ in Equation (\ref{model2f}). CAR priors are widely used in spatial analysis to encourage similar parameter estimates at neighboring locations \citep{Besag:1991}, and in temporal analysis to smooth parameter estimates at adjacent times \citep{Knorr-Held:1998}.

With such priors, we can represent a rich set of dependence structures, including complex seasonality and high-order dependence structures, which may be especially helpful for analyzing temporal patterns across fine time scales. We can also use unique specification and parameters for each mixture weight, allowing us to detect location-specific temporal patterns.

The mixture weights, $p_{b,j}$, are subject to nonnegativity and sum-to-unity constraints; placing autoregressive priors and manipulating them would require special attention. Instead, we transform them into an unconstrained parametrization via the multinomial logit transformation
\begin{equation} \label{logit}
 \pi_{b,r}=\log\left[\frac{p_{b,r}}{1-\sum_{j=1}^{K-1}p_{b,j}}\right],\,\, \,\,\, \,\, r\in \{1,\ldots,K-1\}, \; b \in \{1,\ldots,B\}.
\end{equation} 
We then specify autoregressive priors on the transformed weights $\{\pi_{b,r}\}$. For this application, we apply the CAR priors to capture first-order autocorrelation and daily seasonality. 

We assume that the de-meaned transformed weights from any time period depend most closely on those from four other time periods: immediately before and after (to represent short-term serial dependence), and exactly one day before and after (to capture daily seasonality). We impose the following priors
\begin{equation} \label{shrinkagepriorslag12} 
\begin{aligned}
& \pi_{b,r}|\boldsymbol\pi_{-b,r}  \sim \mbox{N} \left(c_r+\rho_r\left[(\pi_{b-1,r}-c_r)+(\pi_{b+1,r}-c_r)+(\pi_{b-d,r}-c_r)+(\pi_{b+d,r}-c_r)\right],\nu_r^2\right),\\
& c_r \sim N(0,10^4), \;\;\;\;\;\;\;\; \rho_r  \sim \mbox{U} (0,0.25), \;\;\;\;\;\;\;\; \nu_r^2\sim U(0,10^4),
\end{aligned}
\end{equation}
for $r \in \{1, \ldots, K-1\}$ and $b \in \{1,\ldots,B\}$, in which $\boldsymbol\pi_{-b,r}=(\pi_{1,r},\ldots, \pi_{b-1,r},\pi_{b+1,r},\ldots, \pi_{B,r})'$, and $d$ is the number of time periods in a day ($d=12$ in this case). Since every week has the same sequence of spatial densities, we define priors of $\{\pi_{b,r}\}$ circularly in time, such that the last time period is joined with the first time period. In the prior specification of $\{\pi_{b,r}\}$, the CAR parameters $\rho_r$ determine the persistence in the transformed mixture weights over time, while the intercepts $c_r$ determine their mean levels, and the variances $\nu_r^2$ determine the conditional variability. These three parameters are component-specific, and therefore location-specific. For any $\rho_r \in (-0.25, 0.25)$, the joint prior distribution of $[\pi_{1,r},\ldots, \pi_{B,r}]$ is a proper multivariate normal distribution \citep{Besag:1974}; we take the priors of $\rho_r$ to be $\mbox{U} (0,0.25)$ because exploratory data analysis only detected evidence of nonnegative serial dependence.
The priors on $c_r$ and $\nu_r$ are diffuse, reflecting the fact that we have little prior information regarding their values.  
Alternative to this circular definition of mixture weights with symmetric dependence on past and future, one can also specify the marginal distribution of ${\pi_{1,r}}$ and let each $\pi_{b,r}$ depend only on its past. In either setting, we can represent a wide range of complex temporal patterns. 

This approach can be extended if additional covariates are available. 
For example, if temporal covariates $\boldsymbol{x}_t$ are available (e.g., temperature and precipitation), 
the CAR structure may be applied to the covariate-adjusted weights $\pi_{t,r}-c_{t,r}-\bm{a}_r'\bm{x}_t$.
One advantage of this specification is that it can differentiate the impact of temporal covariates on 
ambulance demand at different component locations in space. In this application, we found no significant temperature effect and only a minor precipitation effect for a small number of components.

\subsection{Bayesian Estimation} \label{compu}

We apply Bayesian estimation, largely following \cite{Richardson:1997} and \cite{Stephens:2000} in our choices of prior distributions and hyperparameters. \cite{Richardson:1997} define a set of independent, weakly informative and hierarchical priors conjugate to univariate Gaussian mixture models, which \cite{Stephens:2000} extends to the multivariate case. We extend to incorporate time-varying mixture weights, and instead of imposing independent Dirichlet priors on $\{p_{b,j}\}$, we impose CAR priors on $\pi_{b,r}$ as in Equation (\ref{shrinkagepriorslag12}). For all other parameters, we have for $j\in \{1, \ldots, K\}$ and $t \in \{1,\ldots,T\}$,
\begin{equation}\label{priors}
\begin{aligned} \boldsymbol\mu_j \sim \mbox{Normal} (\boldsymbol\xi, \boldsymbol\kappa^{-1}),  \;\;\;\;\;\;\; \boldsymbol\Sigma^{-1}_j | \boldsymbol\beta & \sim \mbox{Wishart} (2\alpha, (2\boldsymbol\beta)^{-1}),\\
\boldsymbol\beta & \sim \mbox{Wishart} (2g, (2\boldsymbol h)^{-1}),\end{aligned}
\end{equation}
\noindent  in which we set $\alpha = 3$, $g= 1$ and
\begin{equation*}
\boldsymbol\xi= \left[ \begin{array}{c} \xi_1\\ \xi_2 \end{array} \right], \,\,\,\,\,\, \,\,\,\,\,\,\,\,\, \boldsymbol\kappa=\left[ \begin{array}{cc} \frac{1}{R^2_1} & 0 \\ 0 & \frac{1}{R_2^2} \end{array} \right], \,\,\,\,\,\, \,\,\,\,\,\,\,\,\,  \boldsymbol h = \left[ \begin{array}{cc} \frac{10}{R^2_1} & 0 \\ 0 & \frac{10}{R_2^2} \end{array} \right], 
\end{equation*}
\noindent in which $\xi_1$ and $\xi_2$ are the medians of all observations in the first and second spatial dimensions, respectively, and $R_1$ and $R_2$ are the lengths of the ranges of observations in the first and second spatial dimensions, respectively. The prior on each $\boldsymbol\mu_j$ is diffuse, with prior standard deviation in each spatial dimension equal to the length of the range of the observations in that dimension. The inverse covariance matrices $\boldsymbol\Sigma^{-1}_j$ are allowed to vary across $j$, while centering around the common value $E(\boldsymbol\Sigma^{-1}_j|\boldsymbol\beta)=\alpha\boldsymbol\beta^{-1}$. The constant $\alpha$ controls the spread of the priors on $\boldsymbol\Sigma^{-1}_j$; this is taken to be 3 as in \cite{Stephens:2000}, yielding a diffuse prior for $\boldsymbol\Sigma^{-1}_j$. The centering matrix $\boldsymbol\beta^{-1}$ is given an even more diffuse prior, since $g$ is taken to be a smaller positive constant. Our choice of $\boldsymbol h$ is the same as \cite{Stephens:2000}. 

We perform estimation via Markov chain Monte Carlo (MCMC). We augment each observation $\boldsymbol{s}_{t,i}$ with its latent component label $z_{t,i}$, simulating a Markov chain with limiting distribution equal to the joint posterior distribution of $\{z_{t,i}\}$, $\{\boldsymbol\mu_j\}$, $\boldsymbol\beta$, $\{\boldsymbol\Sigma_j\}, \{\pi_{b,r}\}, \{c_r\}, \{\rho_r\}$, and $\{\nu_r\}$. After initializing all parameters by drawing from their respective priors, we update $\{z_{t,i}\}, \{\boldsymbol \mu_j\}, \boldsymbol \beta$ and $\{\boldsymbol \Sigma_j\}$ by their closed-form full conditional distributions, and update $\{\pi_{b,r}\}, \{c_r\}, \{\rho_r\}$ and $\{\nu_r\}$ via random-walk Metropolis-Hastings.

After estimation, we numerically normalize $f_t(\cdot)$ for each $t$ with respect to Toronto's boundary to obtain final  density estimates. As a result, we predict outside of Toronto's boundary with probability zero, and the density within the boundary is elevated proportionally for each $t$. Note that we did not impose this boundary during estimation, even though it might have been advantageous to do so. There is a relatively high density of observations along the southern boundary near Lake Ontario, and truncating the spatial densities at the boundary would encourage mixture components that are close to the lake to move beyond the lake or take on higher weights to better describe these observations. However, normalizing the spatial densities to the boundary is computationally intensive, requiring numerical integration for every Metropolis-Hastings proposal, for every period, and closed-form full-conditionals for $\boldsymbol{\mu}_j$ and $\boldsymbol{\Sigma}_j$ would no longer exist.


\section{Estimating the Number of Components} \label{nocomponents}

We assumed a fixed number of mixture components in developing the proposed model in Section \ref{3models}; in this section we estimate a variable number of components. 
Allowing the number of components to vary typically improves the mixing (efficiency) of the MCMC computational method, by allowing the Markov chain to escape local modes more quickly. We adapt the birth-and-death MCMC (BDMCMC) computational method from \cite{Stephens:2000} to a spatio-temporal setting. 

Each iteration of Stephens' BDMCMC is a two-stage process. In the first stage, new components are ``born'' or existing ones ``die'' in a continuous time framework. Parameters of new-born components are sampled from their respective priors. Components die at a rate so as to maintain sampling stationarity; they die according to their relative implausibility as computed from the likelihood of observations and priors. After each birth or death, the mixture weights are scaled proportionally to maintain sum-to-one invariance within each time period. After a fixed duration of births and deaths, in the second stage,  the number of components are fixed and distributional parameters and mixture weights of the components are updated using full conditionals or Metropolis-Hastings. 

We can generalize Stephens' BDMCMC to incorporate time-varying mixture weights in a straightforward way, by maintaining the same number of components across different time periods within each iteration. Since the birth and death process applies in the same way to all time periods, it is easy to show that stationarity holds for this generalized sampling method.
Following \cite{Stephens:2000}, we assume a truncated Poisson prior on the number of components $K$, i.e., $P(K) \propto \tau^K/K!, \,\, K\in \{1,\ldots, K_{max}\}$ for some fixed $\tau$ and $K_{max}$. All other priors and hyperparameters are as specified in (\ref{shrinkagepriorslag12}) and (\ref{priors}), and the spatial density function at each time is as in Equation (\ref{model2f}).


\section{Estimating Ambulance Demand}\label{emp}

We fit the full Gaussian mixture model with seasonality constraints and CAR priors (Section \ref{3models}) on the Toronto EMS data from February 2007. First, we use a fixed number of 15 components. We found 15 components to be large enough to capture a wide range of residential, business and transportation regions in Toronto, yet small enough for computational ease given the large number of observations. We then fit the model again with a variable number of components (Section \ref{nocomponents}). Given the large amount of data and the complexity of spatial-temporal methods, imposing a vague prior on the number of components would result in an unfeasibly large number of mixture components, and leads to overfitting. We therefore set the \textit{a priori} maximum number of components $K_{max} = 50$ and chose two small values for the prior mean of the number of components $\tau$. These prior choices lead to posterior average numbers of components of 19 or 24 (with posterior standard deviations of 3.1 and 4.6, respectively).  

Each MCMC algorithm is run for $50,\!000$ iterations, with the first $25,\!000$ iterations discarded as burn-in. 
We compute the effective sample sizes and Gelman-Rubin diagnostics of the minimum and maximum of component means and variances along each spatial dimension. In a typical simulation, the mean parameters have effective sample size averaged 2,606 and Gelman-Rubin below 1.05, and for covariance parameters, 6,065 and 1.09, respectively. This suggests  burn-in and mixing may be sufficient. We focus on the minimum and maximum of these parameters instead of relying on component labels because any mixture models may encounter the label switching problem, in which the labeling of component parameters can permute while yielding the same posterior distribution. However, this label switching problem does not affect estimation of ambulance demand in time and space, because we are interested in the entire posterior distribution, instead of inferences on individual mixture parameters. In Section \ref{gof} we also report the estimated MCMC standard errors of the performance measures \citep{Flegal:2008}; they are small enough to provide accuracy to 3 significant digits, further suggesting the run length may be satisfactory. 

Using a personal computer, the computation times for the proposed model with 15 components is about 4 seconds per iteration, compared to 7 and 8 seconds using variable numbers of components averaged 19 and 24, respectively.
In practice, estimation using the proposed model only needs to be performed infrequently (at most once a month in this application); density prediction of any future time period can then be immediately calculated as the corresponding density using the most recent parameter estimation results.

\begin{figure}[h!]
	\centering
\includegraphics[width=0.7\textwidth]{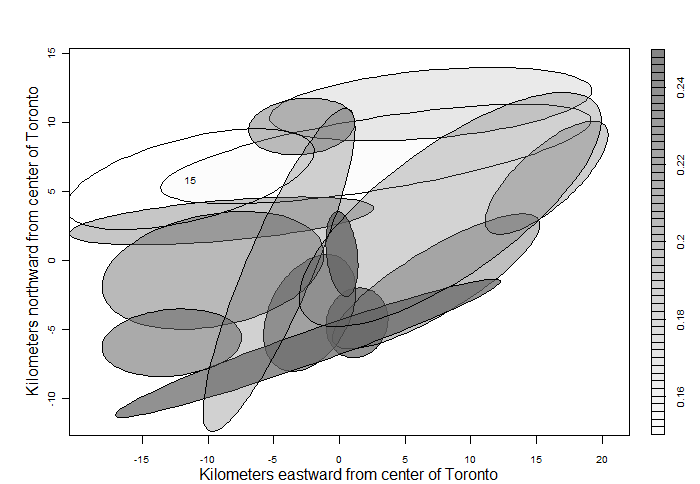}
	\caption{Ellipses at the 90\% level for all components, when fitting the proposed mixture model using 15 components. Each ellipse (except that of the $15$th component) is shaded with the posterior mean of $\rho_r$ for that component. The greater downtown and coastal regions exhibit stronger low-order serial dependence and daily seasonality.
}
	\label{mean_cov}
\end{figure}

Figure \ref{mean_cov} presents results from fitting using 15 components.  It shows the ellipses at the 90\% level associated with each of the 15 components, using the parameter values from the 50,000th iteration of the Markov chain. The ellipse for each component is shaded by the posterior mean of $\rho_r$ for that component, except for the $15$th component because $r\in \{1,\ldots,14\}$. Components at the denser greater downtown and coastal regions of Toronto have the highest estimates of $\rho_r$; these regions exhibit the strongest low-order serial dependence and daily seasonality. The proposed model is able to easily differentiate temporal patterns and dynamics at different locations.

\begin{figure}[h!]
	\centering
\includegraphics[width=1.0\textwidth]{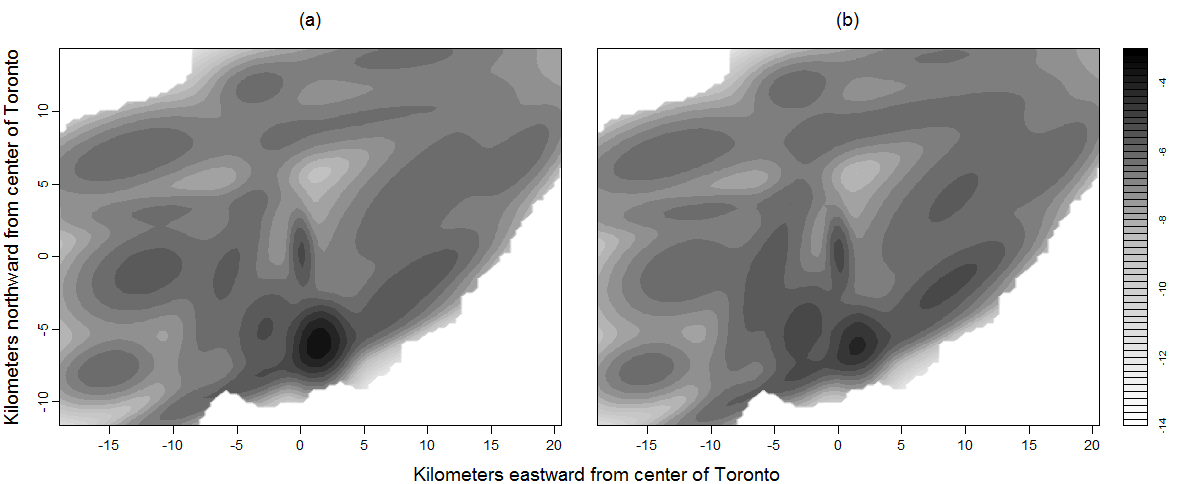}
	\caption{Fitting proposed mixture model using 15 components: (a) posterior log spatial density for Wednesday 2-4pm (demand concentrated at downtown during the day); (b) posterior log spatial density for Wednesday 2-4am (demand more spread out during the night).
}
	\label{fitshrink}
\end{figure}

\begin{figure}[h!]
	\centering
\includegraphics[width=1.0\textwidth]{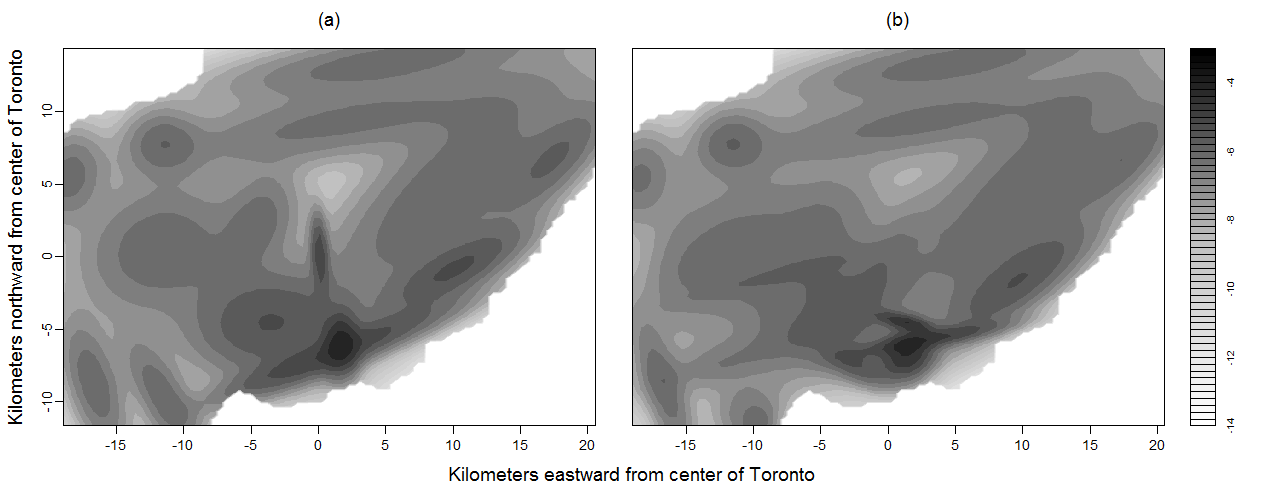}
	\caption{Fitting proposed mixture model using variable number of components: (a) posterior log spatial density for Wednesday 2-4am (night) using an average of 19 components; (b) that using an average of 24 components.
}
	\label{variableplot}
\end{figure}

Figure \ref{fitshrink} show posterior log spatial densities at Wednesday around midday and midnight as computed by the proposed mixture model with 15 components and averaged across the last $25,000$ Monte Carlo samples. Note that the demand is 
concentrated at the heart of downtown during working hours in the day, but is more 
dispersed
throughout Toronto during the night. Figure \ref{variableplot} show the posterior log spatial densities using variable numbers of components around Wednesday midnight; these spatial densities are similar to that using 15 components (shown in Figure \ref{fitshrink}(b)).

\section{Model Performance and Validation} \label{perf}
We evaluate the performance and validity of the proposed models in several ways. For performance, we attempt to predict ambulance demand densities on two sets of test data (March 2007 and February 2008). To do this using mixture models, we train the models on data from February 2007, and use the resulting density estimates to predict for both sets of test data. We introduce in Section \ref{compmtd} two methods for comparisons. 
We compare the statistical predictive accuracies for all methods in Section \ref{gof}. In Section \ref{resptime} we then put these predictive accuracies in the context of EMS operations. We verify the validity of the method in Section \ref{validity}.

\subsection{Comparison Methods} \label{compmtd}
We compare the proposed mixture models to a current industry practice, and to a proposed extension of the industry practice that uses kernel density estimation (KDE). Historically, Toronto EMS employed an averaging method based on a discretized spatial and temporal domains. The demand forecast at a spatial cell in a particular time period is the average of four corresponding realized demand counts for the past four years (from the same location, week of the year, day of the week, and hour of the day). Each spatial cell is 1 km by 1 km. A similar practice described in \cite{Setzler:2009}, the MEDIC method, uses the average of up to twenty corresponding historical demands in the preceding four weeks, for the past five years. These industry practices capture, to various extents, yearly and weekly seasonalities present in EMS demand.

We implement the MEDIC method as far as we have historic data available. Since we focus on predicting the demand density, we normalize demand volumes at any place by the total demand for the time period. For any 2-hour period in March 2007, we average the corresponding demand densities in the preceding four weeks. For any 2-hour period in February 2008, we average the corresponding demand densities in the preceding four weeks in 2008 and those same four weeks in 2007. For example, to forecast the demand density for 8-10am on the second Monday of February 2008, we average the demand densities at 8-10am in the first Monday of February 2008, the last three Mondays of January 2008, the first Monday of February 2007 and the last three Mondays of January 2007. Note that this means the MEDIC method is trained on at least as much data, which is at least as recent as that used in the mixture models. We adopt the same 1 km by 1 km spatial discretization used by Toronto EMS. 

Since the proposed method is continuous in space, we also propose to extend the MEDIC method to predict continuous demand densities as a second comparison method. The demand density for each 2-hour period is taken to be the KDE for all observations from that period.  Here we use a bivariate normal
kernel function, and bandwidths chosen by cross validation using the predictive accuracy measure in Section \ref{gof}. 
We predict demand densities for March 2007 and February 2008 by averaging past demand densities using the MEDIC rule described above. 
To ensure fair comparisons, we also numerically normalize the predictive densities produced by the two comparison methods with respect to Toronto's boundary.

\begin{figure}[h!]
	\centering
\includegraphics[width=\textwidth]{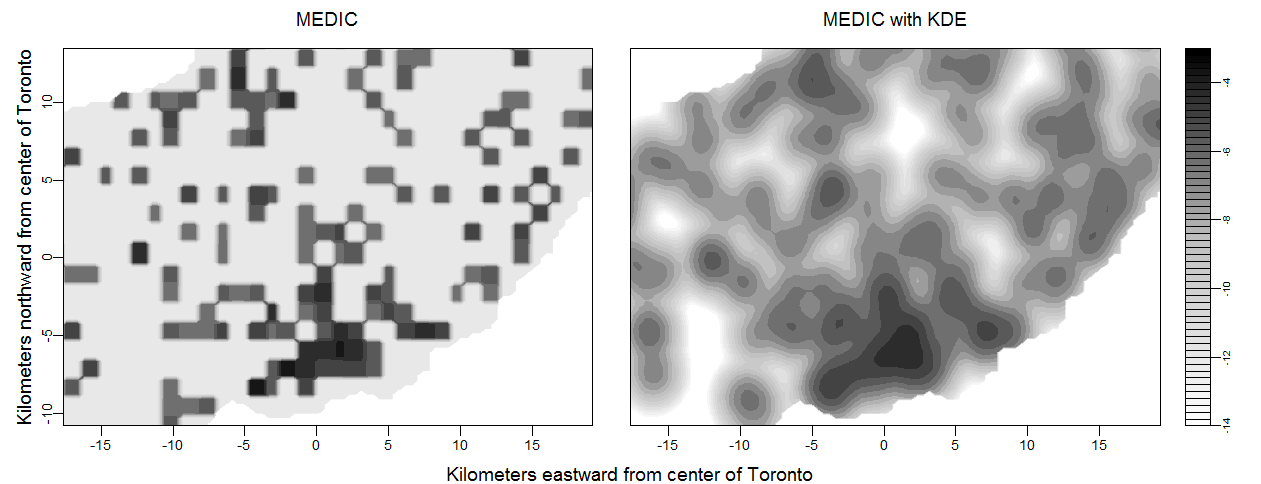}
	\caption{Log predictive densities using two current industry estimation methods for 2-4am (night) on February 6, 2008 (Wednesday). Figure \ref{fitshrink}(b) and Figure \ref{variableplot} show the log predictive densities for the same period using mixture models. Compared to mixture models, estimates from the MEDIC and MEDIC-KDE are less smooth.
}
	\label{medic}
\end{figure}

Figure \ref{medic} shows the log predictive density using these two competing methods for February 6, 2008 (Wednesday) 2-4am. These two densities are comparable with Figure \ref{fitshrink}(b) and \ref{variableplot}, which are the log predictive densities for the same time period estimated from the proposed mixture models. Compared to the proposed model, both the MEDIC and MEDIC-KDE produce less smooth estimates.

\subsection{Statistical Predictive Accuracy} \label{gof}

To measure the predictive accuracy of density estimates obtained from the proposed mixture models, MEDIC, and the proposed MEDIC-KDE, we use the average logarithmic score. First proposed by \cite{Good:1952}, this performance measure is advocated for being a strictly proper scoring rule and its connections with Bayes factor and Bayes information criterion \citep{Gneiting:2007}. We define
\begin{equation}\label{gofeqn}
\mbox{PA}(\{\tilde{\boldsymbol{s}}_{t,i}\})=\frac{1}{\sum_{t=1}^T n_t}\sum_{t=1}^T \sum_{i = 1}^{n_t}\log \hat{f}_t(\tilde{\boldsymbol s}_{t,i}),
\end{equation}
\noindent in which $\hat{f}_t(\cdot)$ is the density estimate for time $t$ obtained using various methods, and $\tilde{\boldsymbol s}_{t,i}$ denotes observations from the test data (March 2007 or February 2008). 
For the proposed mixture models, we use the Monte Carlo estimate of Equation (\ref{gofeqn})
\begin{equation*}
 \mbox{PA}_{mix}(\{\tilde{\boldsymbol s}_{t,i}\})=\frac{1}{M}\sum_{m=1}^{M}\left[\frac{1}{\sum_{t=1}^T n_t}\sum_{t=1}^T \sum_{i = 1}^{n_t}\log \hat{f}_t(\tilde{\boldsymbol s}_{t,i} | \boldsymbol\theta^{(m)})\right],\\
\end{equation*}
in which $\boldsymbol\theta^{(m)}$ represents the $m$th-iteration posterior parameter estimates generated from the training data, for $m\in \{1,\ldots, M\}$ and some large $M$. 

The predictive accuracies of various methods for two test data sets (March 2007 and February 2008) are shown in Table \ref{gofmaintable}. The predictive accuracies for Gaussian mixture models are presented with their 95\% consistent, nonoverlapping batch means confidence intervals \citep[see][]{Jones:2006}, which reflect the accuracy of the MCMC estimates. Here, a less negative predictive accuracy indicates better performance. The proposed mixture models outperform the two current industry methods. Allowing for a variable number of components improves the predictive accuracy slightly, but the rate of improvement diminishes as the average number of components grows. Given that the computational expense almost doubles to obtain these modest improvements, we conclude that using a fixed number of 15 components is largely sufficient in this application.

\begin{table}[h!]
\centering
\resizebox{15cm}{2cm}{ 
\begin{tabular}{|L{2cm}| L{5.5cm}|R{3cm}|R{3cm}|}
\hline
\multicolumn{2}{|C{7.5cm}|}{\textbf{Estimation method}}  & \multicolumn{1}{C{3cm}|}{\textbf{PA for Mar 07}} & \multicolumn{1}{C{3cm}|}{\textbf{PA for Feb 08}} \\ \hline

\multirow{4}{2cm}{Gaussian Mixture} & 15 components ($\S$2) & $-6.1378 \pm 0.0004$ & $-6.1491 \pm 0.0005$ \\ \cline{2-4}

& Variable number of comp ($\S$3): & &\\
& average 19 comp   & $-6.080 \pm 0.002$ & $-6.128 \pm 0.002$ \\ 

& average 24 comp  & $-6.072 \pm 0.003$ & $-6.122 \pm 0.004$ \\ \hline

\multirow{2}{2cm}{Competing Methods} & MEDIC & $-8.31$ & $-7.62$ \\ \cline{2-4}

& MEDIC-KDE  & $-6.87$ & $-6.56$ \\ \hline
\end{tabular}}
\caption{Predictive accuracies of proposed Gaussian mixture models and competing methods on test data of March 2007 and February 2008. The predictive accuracies for mixture models are presented with their 95\% batch means confidence intervals. }
\label{gofmaintable}
\end{table}


\subsection{Operational Predictive Accuracy} \label{resptime}
In this section, we quantify the advantage of the proposed model over the industry practice. We show that the proposed model gives much more accurate forecasts of the industry's operational performance measure. 
The standard EMS operational performance measure is the fraction of events with response times below various thresholds (e.g., 60\% responded within 4 minutes). 
Obtaining an accurate forecast of this performance is of paramount importance because many aspects of the industry's strategic management aim to optimize this performance. 
Accuracy in estimating this performance depends crucially on the accuracy of spatio-temporal demand density estimates. 

For each of the three methods of interest, we have a set of 2-hour demand density estimates for March 2007 and February 2008. Using density estimates generated by method $\mathscr{M}$ for time period $t$,  we predict the operational performance by computing the proportions of demand, $\mathscr{P}_{\mathscr{M}, t} (r)$, reachable within response time threshold $r$ from any of the 44 ambulance bases in Toronto (see Figure \ref{resptime}(a)). 
To do so, we first discretize Toronto into a fine spatial grid and outline the regions that can be covered within any response time threshold. We then numerically integrate within these regions the demand density estimates from $\mathscr{M}$, for each $t$ and $r$, to obtain $\mathscr{P}_{\mathscr{M}, t}(r)$. We also compute the realized performances using the test data, $\mathscr{P}_{test, t} (r)$. 
For simplicity, we assume ambulances always travel at the median speed of Toronto EMS trips, 46.44 km / hour. We also use the  $L_1$ (Manhattan) distance between any base and any location. We consider response time thresholds ranging from 60 seconds to 300 seconds at 10-second intervals. 

\begin{figure}[h!]
	\centering
\includegraphics[width=\textwidth]{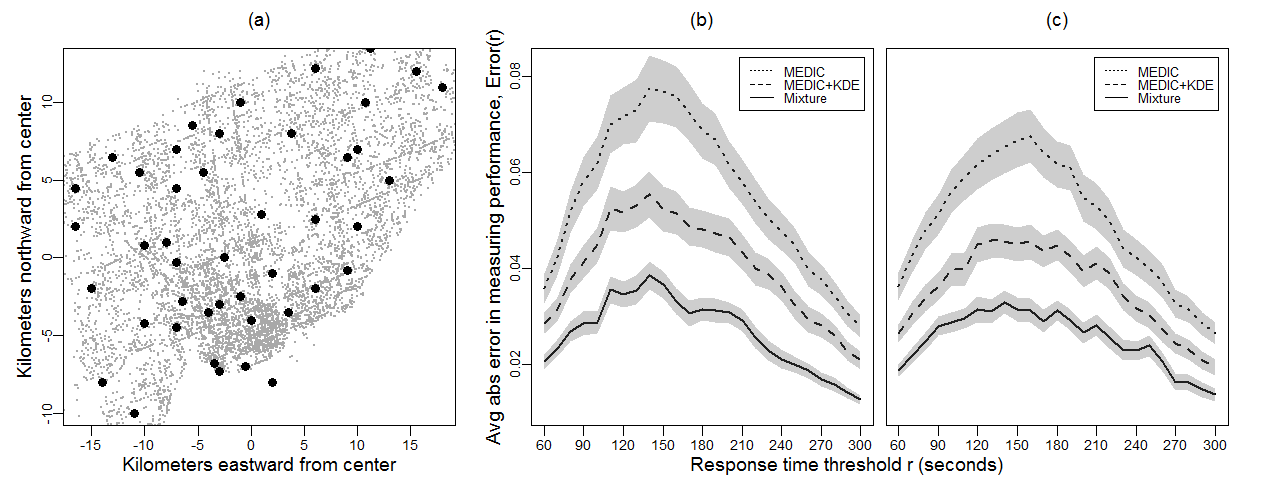}
	\caption{(a) all 44 ambulance bases in Toronto; (b) and (c) average absolute error in measuring operational performance made by the proposed mixture model (15 components), MEDIC, and MEDIC-KDE, using test data from March 2007 and February 2008, respectively (with 95\% point-wise confidence intervals in gray). The proposed mixture model outperforms the competing methods. 
}
	\label{resptimeplot}
\end{figure}

We compute the average absolute error in predicting operational performance made by each method under various response time thresholds, as compared to the truth. We define $\mbox{Error}(\mathscr{M},r) = \frac{1}{T}\sum_{t=1}^T |\mathscr{P}_{\mathscr{M},t} (r) - \mathscr{P}_{test,t} (r)|$. In Figure \ref{resptimeplot} (b) and (c), we show $\mbox{Error}(\mathscr{M},r)$ against $r$ for each method $\mathscr{M}$ (mixture model with 15 components, MEDIC, and MEDIC-KDE), using test data from March 2007 and February 2008, respectively. The $95\%$ point-wise confidence bands for $\mbox{Error}(\mathscr{M},r)$ are shown in gray; these bands indicate interval estimates for the average absolute errors for each $\mathscr{M}$ and $r$ given a  series of errors.
We find that the proposed method predicts the operational performance much more accurately, given the same set of operational assumptions about base locations, speed and distance. It reduces error by as much as two-thirds compared to the MEDIC method, despite sometimes using less recent training data. We expect similar orderings of the three methods under different sets of operational strategies.

\subsection{Model Validation} \label{validity}
\begin{figure}[h!]
	\centering
\includegraphics[width=\textwidth]{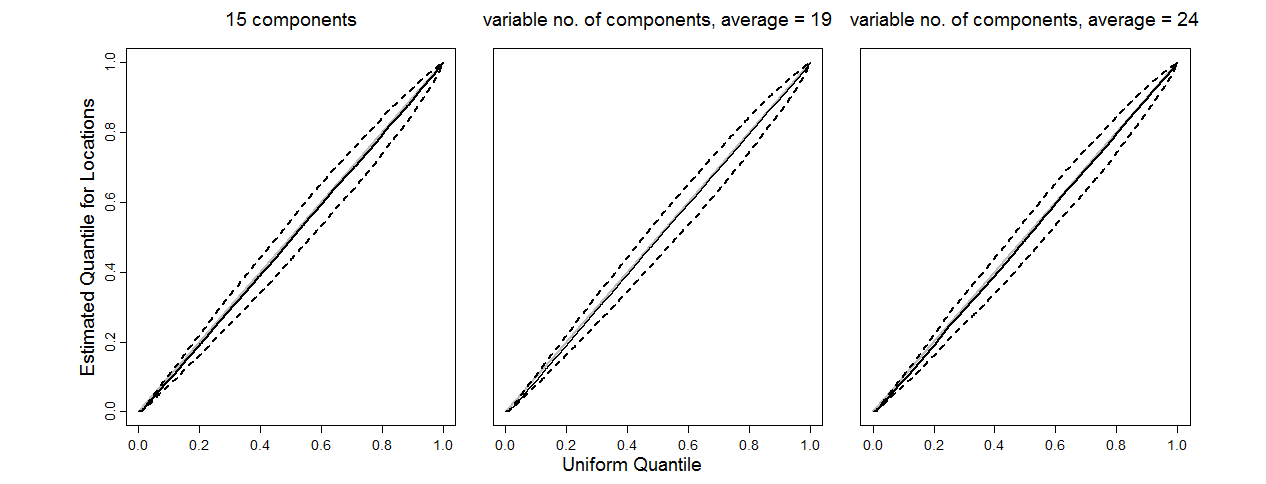}
	\caption{Posterior Q - Q plots for locations using the proposed mixture models, where the solid lines show the posterior mean Q - Q lines and the dash lines are the 95\% posterior intervals. All three plots show that the models are adequate and appropriate.
}
	\label{valplot}
\end{figure}

We assess the goodness-of-fit of the proposed models and the validity of the NHPP assumption. We use the model checking approach by \cite{Taddy:2012}, where each marginal of the point event data is transformed into quantities that are assumed to be uniformly distributed, and compared to the true uniform distribution graphically. In particular, we have assumed that the point process follows a NHPP with time-varying intensity $\lambda_t(\boldsymbol{s})=\delta_t f_t(\boldsymbol{s})$. We have posterior estimates of $\{f_t(\boldsymbol{s})\}$ from the proposed mixture models, and since we do not estimate $\delta_t$, we assume that $\delta_t=n_t$, in which $n_t$ is the actual, realized demand counts in the $t$-th period. 
The uncertainties in estimating $\delta_t$ are not our focus as we attempt to validate the proposed model only with respect to the spatial densities; however, the results below were similar when using estimates of $\delta_t$.
Point locations along the first and the second spatial dimension thus follow one-dimensional NHPP with marginalized intensities of $\lambda_t(\cdot)$, denoted as $\lambda_{1,t}(\cdot)$ and $\lambda_{2,t}(\cdot)$, respectively. We compute the corresponding cumulative intensities $\Lambda_{1,t}(\cdot)$ and $\Lambda_{2,t}(\cdot)$ and sort the observations for each time period into ordered marginals $\{\bar s_{j,1}, \ldots, \bar s_{j,n_t}\}$ for each dimension $j \in \{1,2\}$. If the assumptions are valid and the models have perfect goodness-of-fit, then $\{\Lambda_{j,t}(\bar s_{j,i}): i = 1,\ldots, n_t\}$ for each $t$ and $j \in \{1,2\}$ follows a homogeneous Poisson process with unit rate, and $u_{i,j,t}=1-\exp\{-(\Lambda_{j,t}(\bar s_{j,i})-\Lambda_{j,t}(\bar s_{j,i-1})\}$ for $i \in \{1,\ldots n_t\}$, $j \in \{1,2\}$ and $t\in \{1,\ldots, T\}$ are i.i.d uniform random variables on (0,1). We compare the $u_{i,j,t}$ samples obtained from the models with the uniform distribution via quantile-quantile (Q-Q) plots. We have a set of $\{u_{i,j,t}\}$ for each set of posterior parameter estimates. In Figure \ref{valplot} we show the mean Q-Q line and the 95\% point-wise intervals reflecting the uncertainty in MCMC sampling. All three plots indicate high goodness-of-fit, whether we are using a fixed or a variable number of components.

\section{Conclusions} \label{conclusions}
Estimating ambulance demand accurately at fine temporal and spatial resolutions is critical to ambulance planning. The current industry practice and other earlier methods are often simple and do not give accurate estimates. We provide a much-needed method to model spatio-temporal ambulance demand in Toronto using finite mixture models, capturing the complex temporal patterns and dynamics in this large-scale dataset. We demonstrate that the proposed method predicts the EMS operational performance much more accurately, reducing error by as much as two-thirds compared to an industry practice. Many management decisions seek to optimize this estimated operational performance; the proposed method predicts this optimization objective with more accuracy, leading to more confidence in optimization.

We have also developed a set of easily generalizable tools suitable to analyze a wide range of spatio-temporal point process applications. We jointly estimate mixture component distributions over time to promote efficient learning of spatial structures, and describe spatial and temporal characteristics using mixture weights. This approach can be applied to various settings in which particular spatial aspects of the point process are time invariant, or data are too sparse at the desired temporal granularity to describe spatial structures accurately. The evolution of mixture weights provides a flexible and simple framework to explore complex temporal patterns, dynamics, and their interactions with space in a spatio-temporal point process. In this application, we capture seasonalities by constraining the mixture weights, and represent any location-specific dependence structure by imposing CAR priors on the mixture weights. We have also shown that estimation may be implemented with a variable number of components. The proposed method is parsimonious, flexible, straightforward to implement, and computationally-feasible for large-scale datasets.

We propose a method that utilizes the same data as the current industry methods, and does not require any additional data collection. Future work can investigate the use of additional covariates, such as weather, special events, population and demographic variables, in addition to historical data. A further challenge is to collect and make use of data on population and demographic shifts across fine time scales, e.g., hourly.
Additionally, a computationally-feasible way of incorporating the boundary of Toronto and accounting for the high concentration of observations near the boundary would be an important contribution.


\baselineskip=12pt

 \section*{Acknowledgments}
The authors sincerely thank Toronto EMS for sharing their data. This research was partially supported by NSF Grant CMMI-0926814, NSF Grant DMS-1209103 and NSF Grant DMS-1007478.


\setlength{\gnat}{12pt} 
\baselineskip=\gnat

\bibliographystyle{ECA_jasa}
\bibliography{bibliography}		

\end{document}